\documentclass[preprint,3p,times]{elsarticle}

\graphicspath{ {figures/} } 
\usepackage[outdir=./figures/]{epstopdf}

\usepackage{amssymb}
\usepackage{mathrsfs}
\usepackage{txfonts}
\usepackage{amsthm}

\usepackage{color}
\PassOptionsToPackage{linktocpage}{hyperref}
\usepackage[hyperindex,breaklinks]{hyperref}

\newcommand*{\ANL}{Argonne National Laboratory, Argonne, Illinois 60439}

\newcommand*{\ASU}{Arizona State University, Tempe, Arizona 85287-1504}

\newcommand*{\CSUDH}{California State University, Dominguez Hills, Carson, CA 90747}

\newcommand*{\CANISIUS}{Canisius College, Buffalo, NY}

\newcommand*{\CMU}{Carnegie Mellon University, Pittsburgh, Pennsylvania 15213}

\newcommand*{\CUA}{Catholic University of America, Washington, D.C. 20064}

\newcommand*{\SACLAY}{IRFU, CEA, Universit'e Paris-Saclay, F-91191 Gif-sur-Yvette, France}

\newcommand*{\CNU}{Christopher Newport University, Newport News, Virginia 23606}

\newcommand*{\UCONN}{University of Connecticut, Storrs, Connecticut 06269}

\newcommand*{\FU}{Fairfield University, Fairfield CT 06824}

\newcommand*{\FERRARAU}{Universita' di Ferrara , 44121 Ferrara, Italy}

\newcommand*{\FIU}{Florida International University, Miami, Florida 33199}

\newcommand*{\FSU}{Florida State University, Tallahassee, Florida 32306}

\newcommand*{\Genova}{Universit$\grave{a}$ di Genova, 16146 Genova, Italy}

\newcommand*{\GWUI}{The George Washington University, Washington, DC 20052}

\newcommand*{\ISU}{Idaho State University, Pocatello, Idaho 83209}

\newcommand*{\INFNFE}{INFN, Sezione di Ferrara, 44100 Ferrara, Italy}

\newcommand*{\INFNFR}{INFN, Laboratori Nazionali di Frascati, 00044 Frascati, Italy}

\newcommand*{\INFNGE}{INFN, Sezione di Genova, 16146 Genova, Italy}

\newcommand*{\INFNRO}{INFN, Sezione di Roma Tor Vergata, 00133 Rome, Italy}

\newcommand*{\INFNTUR}{INFN, Sezione di Torino, 10125 Torino, Italy}

\newcommand*{\ORSAY}{Institut de Physique Nucl\'eaire, CNRS/IN2P3 and Universit\'e Paris Sud, Orsay, France}

\newcommand*{\ITEP}{Institute of Theoretical and Experimental Physics, Moscow, 117259, Russia}

\newcommand*{\JMU}{James Madison University, Harrisonburg, Virginia 22807}

\newcommand*{\KNU}{Kyungpook National University, Daegu 41566, Republic of Korea}

\newcommand*{\MISS}{Mississippi State University, Mississippi State, MS 39762-5167}

\newcommand*{\UNH}{University of New Hampshire, Durham, New Hampshire 03824-3568}

\newcommand*{\NSU}{Norfolk State University, Norfolk, Virginia 23504}

\newcommand*{\ODU}{Old Dominion University, Norfolk, Virginia 23529}

\newcommand*{\RPI}{Rensselaer Polytechnic Institute, Troy, New York 12180-3590}

\newcommand*{\ROMAII}{Universita' di Roma Tor Vergata, 00133 Rome Italy}

\newcommand*{\MSU}{Skobeltsyn Institute of Nuclear Physics, Lomonosov Moscow State University, 119234 Moscow, Russia}

\newcommand*{\SCAROLINA}{University of South Carolina, Columbia, South Carolina 29208}

\newcommand*{\TEMPLE}{Temple University,  Philadelphia, PA 19122 }

\newcommand*{\JLAB}{Thomas Jefferson National Accelerator Facility, Newport News, Virginia 23606}

\newcommand*{\UTFSM}{Universidad T\'{e}cnica Federico Santa Mar\'{i}a, Casilla 110-V Valpara\'{i}so, Chile}

\newcommand*{\EDINBURGH}{Edinburgh University, Edinburgh EH9 3JZ, United Kingdom}

\newcommand*{\GLASGOW}{University of Glasgow, Glasgow G12 8QQ, United Kingdom}

\newcommand*{\VT}{Virginia Tech, Blacksburg, Virginia   24061-0435}

\newcommand*{\VIRGINIA}{University of Virginia, Charlottesville, Virginia 22901}

\newcommand*{\WM}{College of William and Mary, Williamsburg, Virginia 23187-8795}

\newcommand*{\YEREVAN}{Yerevan Physics Institute, 375036 Yerevan, Armenia}

\title{Differential Cross Section for $\gamma d \rightarrow \omega d$ using 
CLAS at Jefferson Lab}

\journal{Physics Letters B}

\begin{document}

\begin{frontmatter}

\author[toOU]{T.~Chetry}\ead{tc558111@ohio.edu}
\author[toOU]{K.~Hicks}
\author[toOU]{N.~Compton}
\author[toFIU]{M.~Sargsian}

\author[toFIU]{S.~Adhikari}
\author[toSACLAY]{J.~Ball}
\author[toINFNFE]{I.~Balossino}
\author[toINFNFE]{L. Barion}
\author[toINFNGE]{M.~Battaglieri}
\author[toJLAB,toKNU]{V.~Batourine}
% ...
 
 \address[toOU]{Ohio University, Athens, Ohio 45701}
% ...
\address[toSACLAY]{\SACLAY} 
% ...
\address[toFIU]{\FIU} 
% ...
\address[toINFNFE]{\INFNFE} 
% ...
\address[toINFNGE]{\INFNGE} 
% ...
\address[toKNU]{\KNU} 
% ...
\address[toJLAB]{\JLAB} 
% ...


\begin{abstract}
The cross section for coherent $\omega$-meson photoproduction
off the deuteron has been measured for the first time as a function
of the momentum transfer $t = (P_{\gamma}-P_{\omega})^2$
and photon energy $E_{\gamma}$ using the CLAS detector at the 
Thomas Jefferson National Accelerator Facility. The cross sections are
measured in the energy range $1.4 < E_{\gamma} < 3.4$ GeV. A 
model based on $\omega-N$ rescattering is consistent with the data
at low and intermediate momentum transfer, $|t|$. For 
$2.8 < E_{\gamma} < 3.4$ GeV, the total cross-section of $\omega-N$ 
scattering, based on fits within the framework of the Vector Meson 
Dominance model, is in the range of 30-40 mb.
\end{abstract}


\end{frontmatter}

\twocolumn
\section{Introduction}
\label{sec:intro}
\begin{figure}[hbt]
\centering
\includegraphics[width=0.35\textwidth, keepaspectratio = true]{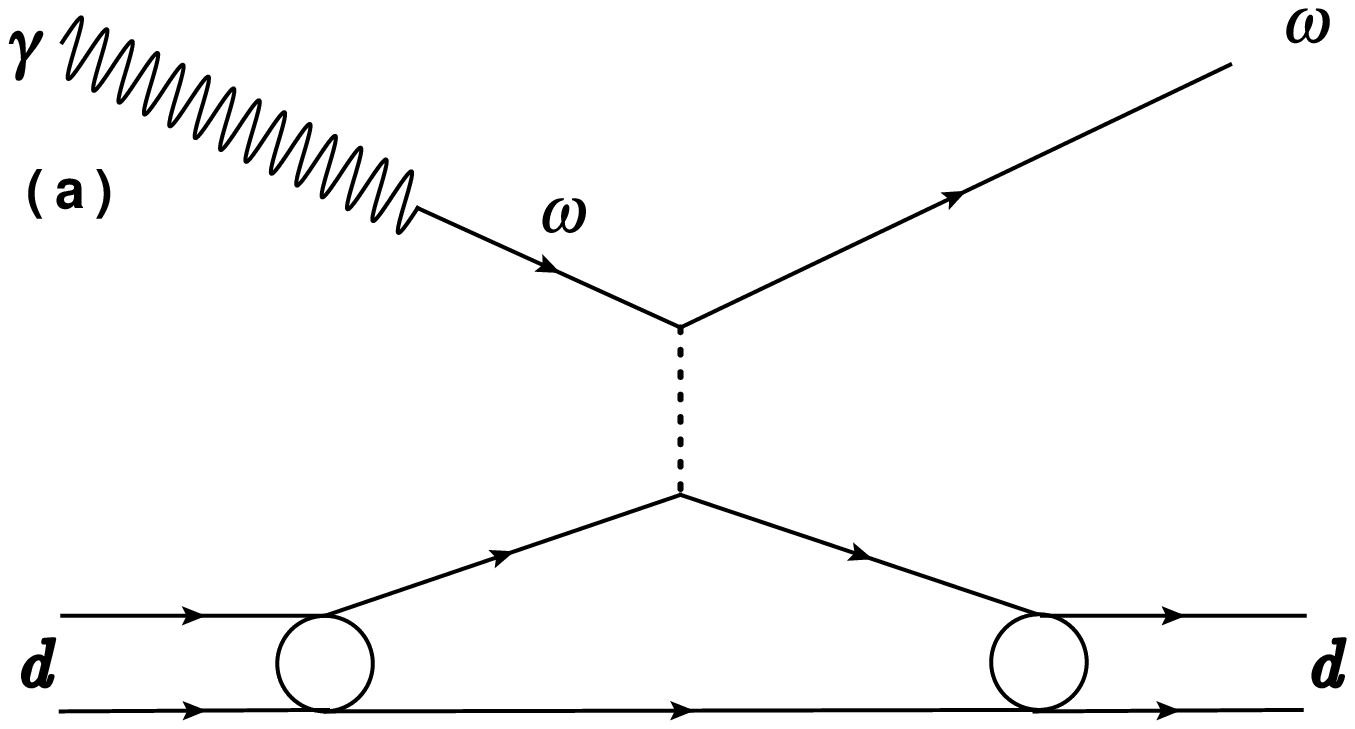}

\includegraphics[width=0.35\textwidth, keepaspectratio = true]{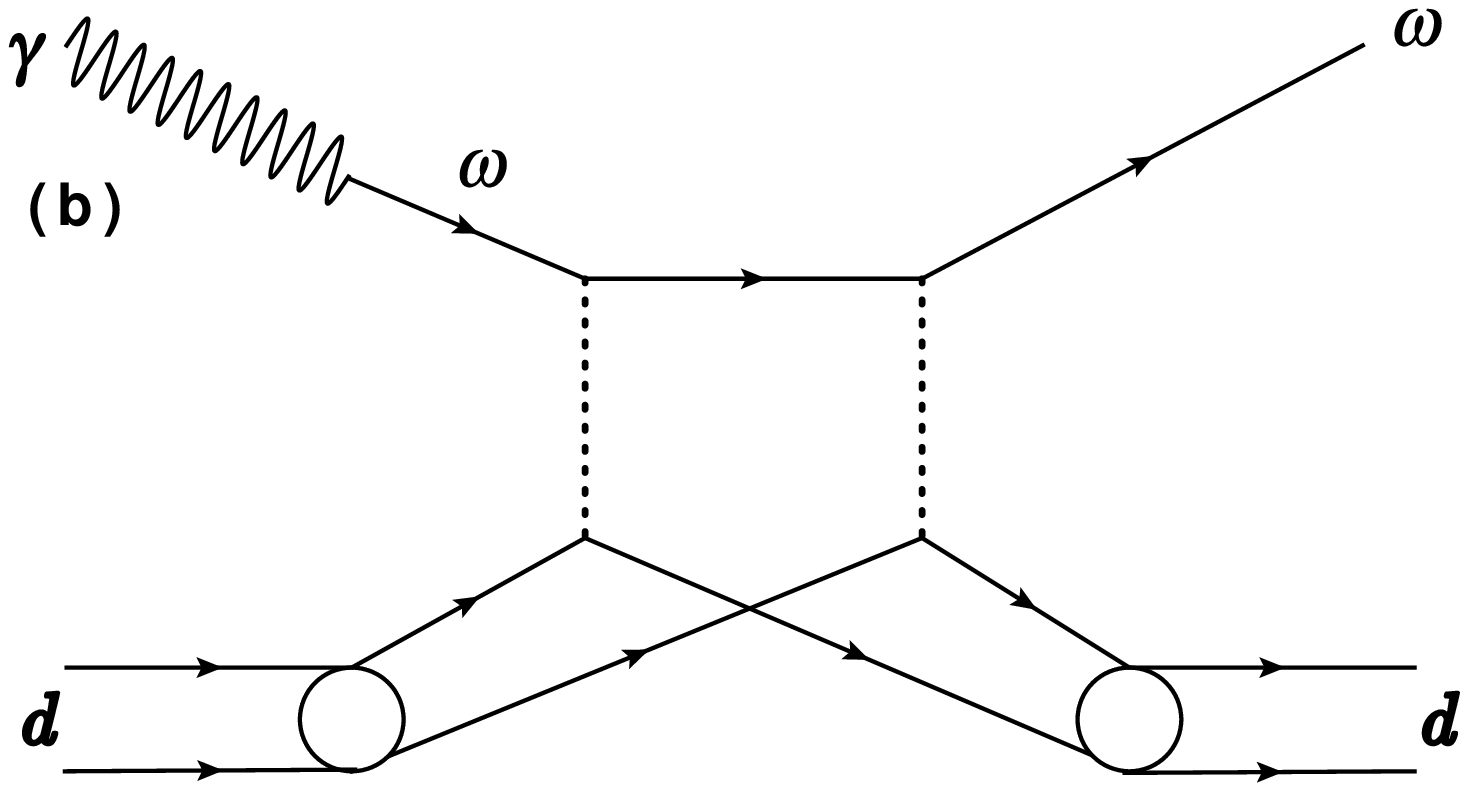}
\caption{\label{fig:scattering} $t$-channel diagrams showing two $\omega$-meson
    photoproduction mechanisms via {\bf(a)} Single and {\bf(b)} Double scattering.
    The photon fluctuates to the $\omega$, which scatters off the nucleon(s). The 
    dashed line represents the exchange of a Pomeron.}
\end{figure}

Vector meson photoproduction off protons at high energies 
is well described \cite{bauer78} theoretically using 
the phenomenological Vector Meson Dominance (VMD) model, in which
the photon fluctuates into a virtual light vector meson (having the 
same quantum numbers as the photon) and then scatters off the target 
\cite{SAKURAI19601}. The VMD model has been very successful at 
predicting vector meson production at high energies. However, at 
photon energies closer to the production threshold, other diagrams, 
such as pseudoscalar meson exchange in the 
$t$-channel, can contribute \cite{OhTitovLee}. This makes the 
reaction dynamics of vector meson photoproduction off proton 
targets more complex near threshold. Additional
complexity near the threshold may come from nucleon resonances
in the $s$-channel.
Coherent $\omega$-meson 
production off the deuteron avoids such complexities. 
Since both the deuteron and the final 
$\omega d$ state are isosinglets, exchange of non-isosinglet 
(e.g. pseudoscalar) mesons cannot contribute. Hence, natural 
parity exchange in the $t$-channel, usually described by Pomeron 
exchange (see Fig.~\ref{fig:scattering}a), is expected to dominate
at low momentum transfer (low $|t|$, where 
$t = (P_{\gamma} - P_{\omega})^2$ and $P_i$ is the four-momentum 
of particle $i$) for vector meson photoproduction off deuterium, 
thus simplifying theoretical interpretations of the data.

At higher momentum transfer ($|t| > 0.5$ GeV$^2/c^2$) secondary 
scattering diagrams, where the $\omega$ is produced off one 
nucleon and scatters from the second, as shown in 
Fig.~\ref{fig:scattering}b, enable both nucleons to remain bound
as a deuteron in the final state \cite{frankfurt97}. These 
diagrams provide an opportunity to extract the $\omega-N$ total 
scattering cross section, $\sigma_{\omega N}$, from comparisons 
of data and calculations. Similar studies were done for coherent 
$\phi$-meson photoproduction from the deuteron \cite{mibe2007,Chang2008},
resulting in the first-ever estimates of the $\phi-N$ total cross
section. Information on these vector meson-nucleon total 
cross sections is virtually impossible to extract cleanly via 
other methods, due to the short lifetimes of these mesons.

Experimental information on $\sigma_{\omega N}$ is of interest currently 
due to progress within lattice QCD, which can now extract
meson-meson scattering phase shifts directly \cite{dudek2015}. The
Hadron Spectroscopy Collaboration \cite{dudek2015} is working
on extracting meson-nucleon scattering phase shifts, 
which are directly related to the total cross sections.
This is a significant advance because it connects QCD calculations
to experimental observables, such as the total cross sections.
Such a direct connection between non-perturbative QCD and experiment 
has been rare until now.
The $\omega$ meson is a particularly good choice for these studies, 
since it decays into three pions about 89\% of the time.
On the lattice, the light quark masses are inputs. 
Lattice results are often shown for pion masses heavier than in 
nature, as the lattice calculations are easier to compute there.
The $\omega$ is thus a stable particle in lattice calculations where 
the pion mass is somewhat higher than its physical value.  
Scattering phase shifts of stable particles are easier to obtain 
on the lattice than for unstable particles.
Hence, measurements of $\sigma_{\omega N}$ are timely and 
can soon be compared with predictions from lattice calculations.

Previous experimental data on coherent $\omega$ photoproduction are 
scarce. Bubble chamber measurements \cite{bauer78} have low statistical
precision. The best data on this reaction are from the Weizmann 
Institute \cite{eisenberg76}, using a photon beam of energy 4.3 GeV
and at $|t|<0.2$ GeV$^2/c^2$, which is too small to see the
effect of double-scattering as shown in Fig.~\ref{fig:scattering}b. 
Data on coherent $\rho$ photoproduction have been measured at higher 
$|t|$ at SLAC \cite{anderson71}, which was used to extract
$\sigma_{\rho N}$. No previous data exist that would allow an 
extraction of $\sigma_{\omega N}$.  

Here, we present data on coherent $\omega$ photoproduction off 
deuterium at photon energies ranging from 1.4 to 3.4 GeV over a wide 
range in the momentum transfer $t$. The $t$-dependence of the cross 
section is measured out to $|t|\sim 2.0$ GeV$^2/c^2$, which is compared 
with theoretical calculations that include the double-scattering 
diagrams, allowing an extraction of the total scattering cross section 
$\sigma_{\omega N}$. This completes the measurement of scattering 
cross sections for the trio of vector mesons ($V=\rho,\omega,\phi$).

\section{Experiment}
\label{sec:expt}
The $g$10 dataset, with unpolarized beam, was collected in the spring of 2004 
using the Continuous Electron Beam Accelerator Facility (CEBAF)  
and the CEBAF Large Acceptance Spectrometer (CLAS) at the Thomas 
Jefferson National Accelerator Facility (Jefferson Lab). CLAS was
designed around six superconducting coils arranged in a toroidal
configuration that produced a field in the azimuthal direction.
The particle detection system consisted of six sets of drift chambers to
determine charged-particle trajectories, gas
Cherenkov counters to identify electrons, scintillator counters for measuring
the time-of-flight (TOF) and electromagnetic calorimeters to detect neutrons
and showering particles such as electrons. These segments
were instrumented individually so that they formed actually independent
magnetic spectrometers with a common target, trigger and data-acquisition
system~\cite{Mecking2003513}.

The $g$10 experiment used a continuous electron beam with incident electron energy,
$E_e = 3.767$ GeV. This beam produced bremsstrahlung photons when passed
through a thin gold radiator \cite{Mecking2003513}. The tagger system 
\cite{Sober2000263} was used to measure the energy of the photons, which
interacted with an unpolarized liquid deuterium target measuring 24 cm 
in length and 4 cm in diameter. The reaction products traversed 
the large drift chambers and timing detectors. The three-momenta 
were reconstructed by the drift chambers and
the particle identification was determined by the arrival time of the products.

The data acquisition trigger required two charged particles detected 
in coincidence with the tagged photon. The time of flight of a particle 
was determined using the scintillator paddles in the start counter 
\cite{Sharabian2006246} that surrounded the target and the 
Time of Flight (ToF) scintillator paddles that surrounded the exterior 
of CLAS \cite{SMITH1999265}. The  charge of the particle was determined 
by its direction of bending in the magnetic field. We used the lower magnetic 
field (torus magnet current set at 2250 A) $g$10 dataset to optimize the 
acceptance for low-momentum in-bending $\pi^-$ \cite{nickg10}.

\section{Data analysis}
\label{sec:dtanlys}
The exclusive reaction $\gamma d \rightarrow \omega d$ was identified 
by detecting a final-state deuteron and two charged pions from 
$\omega \rightarrow \pi^+ \pi^- \pi^0$ decay. The unmeasured 
$\pi^0$ was reconstructed from the missing mass. Charged particles 
were identified from their measured three-momentum and measured 
flight time, using
\begin{equation}
 \delta t=t_{measured} - \frac{d_{path}}{c}\frac{\sqrt{p^2 + m^2}}{p},
\end{equation}
where $d_{path}$ is the reconstructed path length of the particle from the
event vertex to the ToF paddles, $p$ is its momentum, and $m$ is the 
assumed mass. The time difference for a charged particle about 
$\delta t = 0$ was fit as a function of the particle momentum with a 
Gaussian function for several momentum bins. A 3$\sigma$ cut around 
the centroid of $\delta t$ in each momentum bin was used to identify the 
particles in coincidence with a single photon.

The vertex time for each charged particle was compared to the arrival 
time for each photon (from the photon tagger). The photon with the time 
that most closely matched with the vertex time was selected. In order
to remove multiple photons linked to one event, a timing cut of $\pm1$ 
ns was made. This cut helped to avoid ambiguity in selecting the 
``good'' incident photon associated with an event of interest. The events 
rejected by this cut were studied separately and an overall correction, 
$\gamma_{corr}$, of 6.8\% was found. $\gamma_{corr}$ is the ratio of 
the number of rejected photons to the total number of photons 
associated with the final state detected particles for each event.
The measured photon
energy is slightly different from the real photon energy due to a slight
geometrical mismatch, therefore photon energy correction 
(correction factor within $5\%$) explained in \cite{taggerCorrection} was also applied 
to the energy of the selected photons.

With the identity of each scattered particle established, corrections 
were made for the energy lost by each detected charged particle while 
traveling through different materials of the detector \cite{eloss}. 
In addition to the energy loss corrections for the charged particle 
tracks, slight corrections were  also necessary for the momentum of 
each track, due to uncertainties in the magnetic field, by requiring 
four-momentum conservation using the exclusive 
$\gamma d \rightarrow \pi^+\pi^-d$ channel \cite{tayag10}. Cuts were 
also made to remove poorly performing ToF paddles. Events associated 
with beam trips were also removed from the analysis.

In addition to the above, fiducial region cuts were made to remove
events that tracked back to non-physical regions of the detector or
to regions of the detector where the efficiency was low or changing
rapidly. Minimum momentum cuts were used to exclude particles with
low detection efficiency. Events that tracked back to outside the
target region were also removed \cite{tayag10}.

\begin{figure}[hbt]
\centering
\includegraphics[width=0.48\textwidth, keepaspectratio = true]{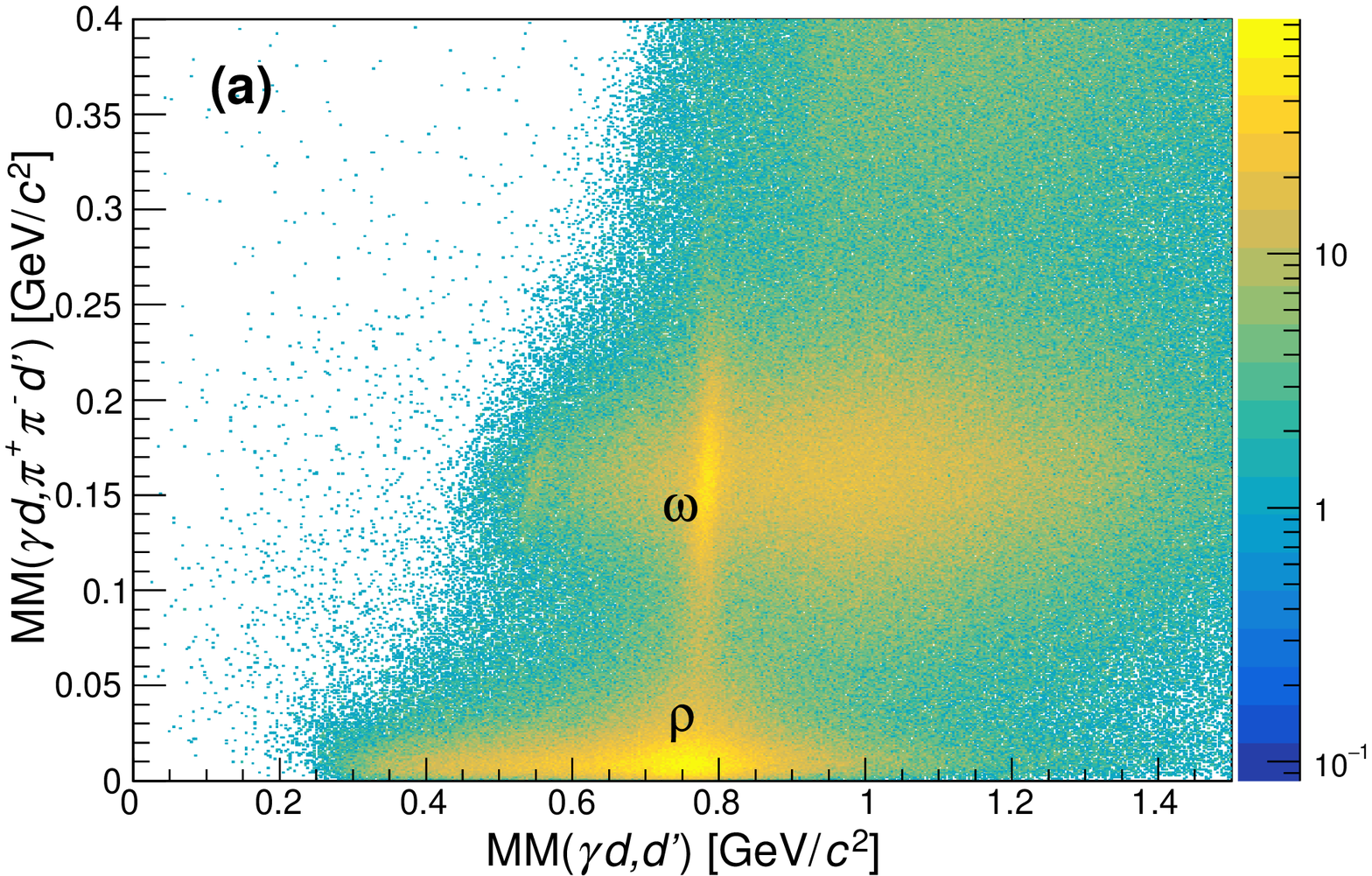}

\includegraphics[width=0.48\textwidth, keepaspectratio = true]{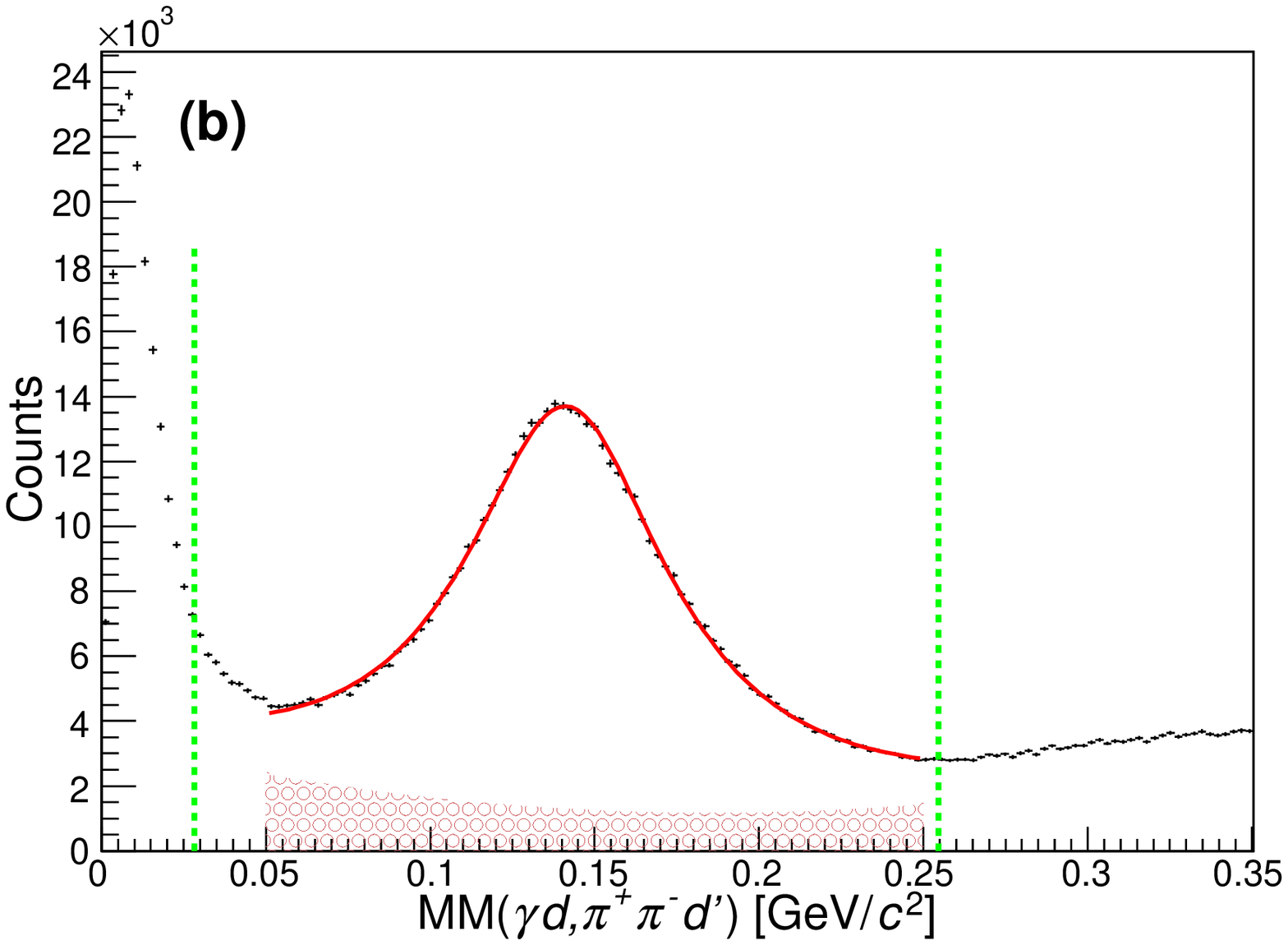}
\caption{\label{fig:mmCut}
  {\bf(a)} The $y$-axis represents the missing mass, $Y$, in the reaction 
  $\gamma d \rightarrow \pi^+ \pi^- d'~Y$ while the $x$-axis shows the 
  distribution of the missing mass, $X$, in the reaction $\gamma d \rightarrow d'~X$.
  {\bf(b)} Shown is the $y$-axis projection of (a). The dashed-green 
  lines represent the position of the missing mass cut made to select 
  $\omega$-events using the fit shown by the solid-red curve. The background 
  is shown by the shaded region.}
\end{figure}

The data analysis for $\gamma d \rightarrow \omega d$ consisted of 
two main steps: two-pion background rejection and $\omega$ yield 
extraction from the multi-meson background. The two-pion 
background is dominated by the $\gamma d \rightarrow \rho d$ 
channel as can be seen from Fig.~\ref{fig:mmCut}a. The majority of
this background was eliminated by requiring that the missing mass 
from the deuteron given by, $MM(\gamma d, d') = \sqrt{(P_d + P_{\gamma} - P_{d'})^2}$, 
equaled $m_{\omega}$, assuming the three pion decay mode for the 
$\omega$. 

A missing $\pi^0$ peak can be seen on top of a smooth 
background in Fig.~\ref{fig:mmCut}b, which was estimated
by employing a Lorentzian for the peak and a second-order 
polynomial for the background. A 3$\sigma$ cut, shown by the 
dashed-green lines in Fig.~\ref{fig:mmCut}b, was applied to select 
exclusive $\omega$ events with some background. 

\begin{figure}[hbt]
\centering
\includegraphics[width=0.48\textwidth, keepaspectratio = true]{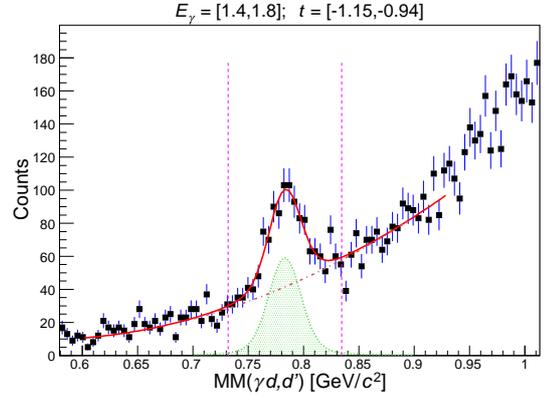}
\caption{\label{fig:yieldE1t5} Yield extraction fit using the missing
    mass distribution for $E_{\gamma} = 1.4-1.8$ GeV and $|t| = 0.94-1.15$ 
    GeV$^2/c^2$. The background is estimated using a polynomial function 
    (dashed-dotted) which, along with the signal (dashed, shaded region), 
    describe the total distribution (solid curve). The yield is the number 
    of events in the shaded region within the dashed-magenta lines.}
\end{figure}

The missing mass spectra were then divided into four $E_{\gamma}$ 
bins in $1.4 < E_{\gamma} < 3.4$ GeV, which were further split 
into different $|t|$ bins within $0.3 < -t < 2.0$ GeV$^2/c^2$.
A total of 25 energy and momentum transfer bins were used to extract the
number of $\omega$-events. The $\omega$-meson yield was 
obtained from a fit to the $MM(\gamma d, d')$ distribution by a
Gaussian-convoluted Lorentzian function, also known as the Voigt profile,
and a background function. The Lorentzian mean and width were fixed 
to 782.65~MeV$/c^2$ and 8.49~MeV$/c^2$ respectively, which correspond to the PDG mass and
width of an $\omega$-meson \cite{PDG}. The Gaussian width, however,
was allowed to vary. A polynomial function of second-order, 
$B(x) = p_1 + p_2x+ p_3x^2$ was chosen to estimate the multi-pion 
background, where $x \equiv MM(\gamma d, d)$, and the $p_i$ are 
fit parameters. The yield was determined by integrating the Voigtian
function. A linear function was used to estimate 
the systematic effect of the choice of the background function. 
An average systematic effect of about 8.6\% was found due to the 
uncertainty of the background subtraction. The fit for one typical 
bin is shown in Fig.~\ref{fig:yieldE1t5}.

Measurement of the differential cross-section required a calculation 
of the CLAS acceptance based upon the geometrical efficiencies of the
detector subsystems. The CLAS acceptance was determined by performing 
a GEANT-based Monte-Carlo simulation. As the acceptance is reaction 
dependent, $\gamma d \rightarrow \omega d \rightarrow \pi^+\pi^- d'(\pi^0)$
events were generated. These events underwent the same event
processing as the data. For each kinematic bin, the acceptance 
was calculated as the ratio of the number of accepted to the
generated events. The simulated distributions were also
fit similar to the data. The average acceptance for this channel
was found to be 8.1\%.

\begin{figure}[hbt]
\centering
\includegraphics[width=0.48\textwidth, keepaspectratio = true]{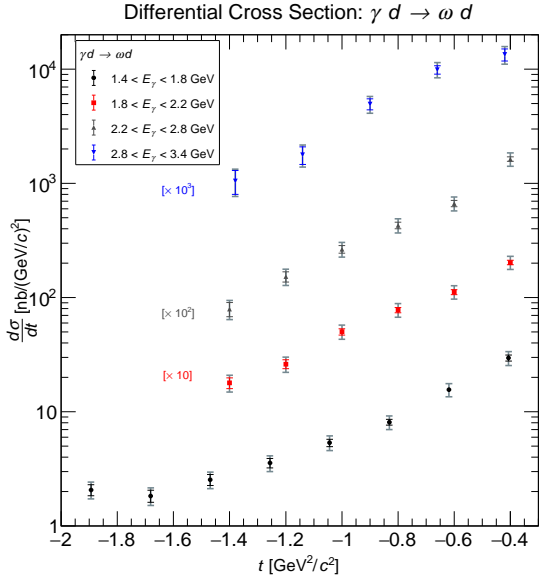}
\caption{\label{fig:plotAlldcs} Differential cross section for 
         $\gamma d \rightarrow \omega d$. The inner error bars represent 
         the statistical uncertainties while the outer error bars include
         systematic uncertainties added in quadrature with the statistical
         uncertainties.}
\end{figure}

The target luminosity was calculated from the incident photon 
flux ($N_{\gamma}$) for the collimated photon beam, target density ($\rho_{T}$), atomic weight 
($M_d = 2.014$ $g/mol$) and length of the target ($l_T$) using the 
relation,
\begin{equation}
 \mathscr{L}(E_{\gamma}) = \frac{\rho_{T} N_A l_{T}}{M_d}N_{\gamma}(E_{\gamma}),
\end{equation}
where $N_A$ is Avogadro's number.
In each energy bin, the differential cross sections
in momentum transfer bins of width $\Delta t$ are calculated 
using the relation,
\begin{equation}
 \frac{d\sigma}{dt} = \frac{Y_D}{\Delta t A \mathscr{L}}\times\frac{\Gamma_{\omega}}
{\Gamma_{\omega\rightarrow\pi^+\pi^-\pi^0}}\times \gamma_{corr}
\label{eq:dcs}
\end{equation}
where $Y_D$ is the yield, $A$ is the detector acceptance, $\mathscr{L}$
is the target luminosity for the photon energy range considered and,
$\Gamma_{\omega\rightarrow\pi^+\pi^-\pi^0}/\Gamma_{\omega}$ is 
the branching ratio. The quantity $\gamma_{corr}$ is the correction factor 
due to the photon selection condition mentioned previously. The statistical
uncertainty on the differential cross section was propagated
from the uncertainties of each quantity in Eq.~\ref{eq:dcs}. The differential 
cross section for $\gamma d \rightarrow \omega d$ is shown in 
Fig.~\ref{fig:plotAlldcs} as a function of $|t|$ for each incident 
photon energy bin.

\begin{table}[hbt]
\centering
\caption{\label{tab:systematicUncertainty} Summary of the g10 systematic 
               effects associated with the $\gamma d \rightarrow \omega d$ 
               channel, estimating a total average point-to-point uncertaity.}
\begin{tabular}{l r}
%
\hline
        Source              & Systematic Uncertainty   \\ \hline
Luminosity/Flux Consistency &         8\%           \\   
Variation of Cuts/Analysis  &         4.4\%           \\   
Yield Extraction            &         8.6\%           \\   
Branching Ratio             &         0.7\%          \\ \hline
Total (Added in quadrature) &         12.5\%          \\ \hline
\end{tabular}
\end{table}

Systematic uncertainties were determined for each part of the 
analysis including the effects of event selection, yield extraction, 
beam normalization, and so on. Table~\ref{tab:systematicUncertainty} 
summarizes the systematic uncertainties calculated in this experiment. 
Due to the large variation of the statistical uncertainty involved in 
this analysis, a point-to-point systematic approach was not realistic. 
Therefore, an estimation of the systematic errors was made by varying 
each cut and taking the average relative difference in the final result 
for each variation. These variations are summarized as different groups
in Table~\ref{tab:systematicUncertainty}.

\section{Results}
\label{sec:rslt}
\begin{figure}[hbt]
\centering
\includegraphics[width=0.48\textwidth, keepaspectratio = true]{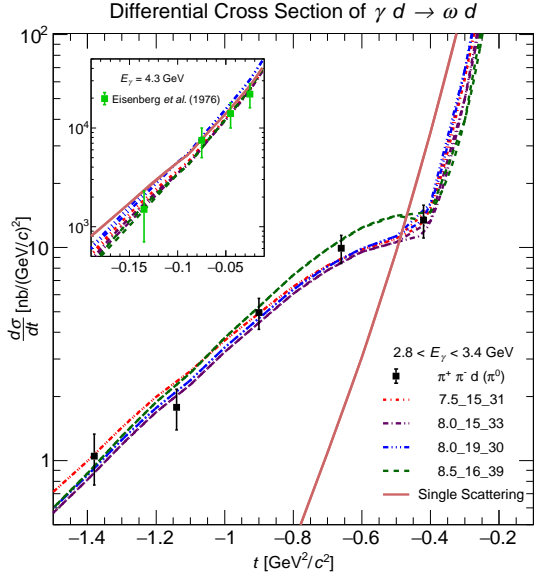}
\caption{\label{fig:dataVsth} Differential cross section of 
        $\gamma d \rightarrow \omega d$
        as a function of $|t|$ for $2.8 < E_{\gamma} < 3.4$ GeV compared to that of 
        a calculation~\cite{misakPrivate} based on \cite{frankfurt97}. Each curve
        corresponds to a specific $b,~\frac{d\sigma}{dt}\big| _{t=0, \gamma N}$ and
        $\sigma_{\omega N}$ value, as listed in Table~\ref{tab:chiSquares}. 
        The legend for each curve is defined respectively for these parameters.
        The solid brown curve represents the contribution of the single scattering
        for input parameters corresponding to that of the red dashed-dotted
        curve. In the inset, the solid points are the results from \cite{eisenberg76} for
        an incident photon energy of $4.3$ GeV.}
\end{figure}

In Fig.~\ref{fig:dataVsth}, the differential cross section for 
$2.8 < E_{\gamma} < 3.4$ GeV is compared with a theoretical calculation
using a rescattering model \cite{frankfurt97}. This is a three-parameter
model allowing us to determine a range of $\sigma_{\omega N}$ by a fit to 
the experimental data. The scattering amplitude of $\gamma N \rightarrow \omega N$
given by
\begin{equation}
 f^{\gamma N \rightarrow \omega N} = \sigma_{\gamma^* \omega}
 (i + \alpha_{\gamma N})e^{\frac{-b_{\gamma N}}{2}t},
\end{equation}
deals with single scattering. A similar equation can be also written for the scattering
amplitude of $\omega N \rightarrow \omega N$ that measures the contribution of the 
rescattering~\cite{frankfurt97}. The quantity $\sigma_{\gamma^* \omega}$ is parametrized using an input parameter
$\frac{d\sigma}{dt}\big| _{t=0, \gamma N}$, which is the differential cross section of
$\gamma N \rightarrow \omega N$ reaction at $t=0$. The initial guess for this parameter was
based on published data on $\gamma p \rightarrow \omega p$~\cite{Dietz2015, marco2002}.
The other two input parameters are $b_{\gamma N}$ or $b_{\omega N}$ 
and $\sigma_{\omega N}$.
At intermediate and higher photon energies, VMD assumes the slope 
factors of the corresponding amplitudes, $b_{\gamma N}$
and $b_{\omega N}$, to be equal. The variables, $\alpha_{\gamma N}$
and $\alpha_{\omega N}$, defined as the ratio of the real to imaginary parts
of the corresponding scattering amplitudes, were kept fixed and equal to a 
phenomenological value of $-0.4$. At intermediate energies, the real part of
the scattering amplitudes for proton and neutron targets are not exactly
the same, but this model omits this difference since $\omega$ production
from $d$ is dominated by isospin averaged amplitudes~\cite{frankfurt97}.
Now, for various slope factors, the strength of single scattering gauged 
by $\frac{d\sigma}{dt}\big| _{t=0, \gamma N}$ is varied. Each set of 
variations was fed to the calculation for various $\sigma_{\omega N}$ 
values as an input parameter and the differential cross section values for a 
particular energy (bin center) were calculated. The outputs were compared
with the data using a $\chi^2$ test. Table~\ref{tab:chiSquares} 
summarizes the values of these parameters that resulted in a reduced 
$\chi^2$ between 0.85 and 1.15. The data is consistent with the 
rescattering model with $30 < \sigma_{\omega N} < 40$ mb in the 
framework of the VMD model.

\begin{table}[hbt]
\centering
\caption{\label{tab:chiSquares} Summary of theory parameters used to 
      compare data for $2.8 < E_{\gamma} < 3.4$ GeV. The parameters 
      shown here are within 15\% of $\chi^2$ = 1.0 (the ideal value).}
\begin{tabular}{c c c c}
\hline
$b_{\gamma N} = b_{\omega N}$ & $\frac{d\sigma}{dt}\big| _{t=0, \gamma N}$ & $\sigma_{\omega N}$ & $\chi ^2/NDF$ \\
$\left[ GeV^{-2}/c^{-2} \right]$ & $[\mu b/ (GeV^{2}/c^2)]$ & $[mb]$ &  \\ \hline
7.5 & 15 & 31 & 1.13 \\ \hline 
8.0 & 14 & 34 & 1.15 \\  
8.0 & 15 & 33 & 1.01 \\ 
8.0 & 16 & 32 & 0.96 \\
8.0 & 17 & 31 & 1.00 \\ 
8.0 & 18 & 30 & 1.15 \\ 
8.0 & 19 & 30 & 0.91 \\
8.0 & 19 & 31 & 0.87 \\
8.0 & 20 & 30 & 1.03 \\ \hline 
8.5 & 16 & 35 & 1.11 \\ 
8.5 & 16 & 39 & 1.00 \\
8.5 & 17 & 34 & 1.05 \\ 
8.5 & 18 & 33 & 1.07 \\ \hline 
9.0 & 19 & 39 & 0.89 \\ 
9.0 & 20 & 38 & 0.87 \\ \hline
\end{tabular}
\end{table}

\section{Conclusion}
\label{sec:conc}
In conclusion, we have presented the first measurement of the differential 
cross sections for coherent $\omega$ photoproduction on the deuteron up to 
$t = -2$ GeV$^2/c^2$ for incident photon energies 1.4 to 3.4 GeV using CLAS
at Jefferson Lab. A model based on rescattering 
is consistent with the data at intermediate and high momentum
transfer. The differential cross section at large $|t|$ 
shows contributions from double scattering. For $2.8 < E_{\gamma} < 3.4$ GeV,
the data is consistent with $\sigma_{\omega N}$ within 30-40 mb. This
range is typical of hadronic cross sections in this energy range. While 
more data would be valuable, this dataset dramatically improves the 
world data on the $\gamma d \rightarrow \omega d$ reaction and opens 
up the possibility for further study of the $\omega-N$ interaction.

\section*{Acknowledgements}
\label{sec:ack}
We are grateful to the staff of the Accelerator and Physics Division at
Jefferson Lab that made this experiment possible and helped us get results
of better quality. This work was supported by: the United Kingdom’s Science 
and Technology Facilities Council (STFC); the Chilean Comisi\`{o}n Nacional 
de Investigaci\`{o}n Cient\`{i}fica y Tecnol\`{o}gica (CONICYT); the Italian
Istituto Nazionale di Fisica Nucleare; the French Centre National de la 
Recherche Scientifique; the French Commissariat \`{a}l’Energie Atomique; 
the U.S. National Science Foundation; and the National Research Foundation 
of Korea. Jefferson Science Associates, LLC, operates the Thomas Jefferson 
National Accelerator Facility for the U.S. Department of Energy under 
Contract No. DE-AC05-06OR23177.




\section{References}
\bibliographystyle{elsarticle-num} 
\bibliography{references}


\end{document}